\documentclass{ijcaArticle}
\usepackage{epstopdf}
\usepackage{amsfonts}
\usepackage{amsmath}
\usepackage{picins}
\ijcaVolume{--}
\ijcaNumber{--}
\ijcaYear{--}
\ijcaMonth{--}

\begin{document}

\title{ A Robust MUSIC Based Scheme for Interference Location in Satellite Systems\\ with Multibeam Antennas} 

\author{ 
   \large Ramoni O. Adeogun \\[-2pt]
   \normalsize School of Engineering and Computer Science \\[-2pt]
    \normalsize Victoria University of Wellington \\[-2pt]
    \normalsize Wellington, New Zealand \\[-2pt]
    \normalsize	ramoni.o.adeogun@ieee.org \\[-2pt]
}

\terms{Satellite Systems, Multibeam Antennas}

\maketitle

\begin{abstract} 
  In this paper, we investigate methods for interference location in satellite communication system using satellite multi-beam antenna with subspace based schemes. A novel MUSIC based approach is proposed for estimating the direction of arrival of the interfering sources. The proposed method provides super resolution and asymptotic maximum likelihood estimates of the direction of arrivals even at low SNR values. Simulations were performed using typical satellite multi-beam antenna configurations and results show that the proposed scheme can effectively estimates the direction of arrival in the azimuth and elevation spectra. Compared to the support vector regression method, the proposed approach offer improved estimation accuracy at low SNR values.
\end{abstract}

\section{Introduction}

Geosynchronous satellite systems are open communication systems that typical suffer from authorized and unauthorized interference or operation. These radio frequency interference causes a degradation in the normal traffic performance of the satellite system \cite{BaiHu2006} and the location of the interfering sources are difficult to identify due to the numerous potential locations for interfering transmitting stations. Another possible reason for the difficulty is because most Earth to satellite transmitting stations direct their transmit powers away from the Earth surface making the detection of terrestrial station difficult except for short range communications \cite{William1989}.  Estimation of the location of interference radiating source from signal received at the satellite is therefore necessary to overcome this difficulty. 

Common methods for determining interference locations include location determination using low-altitude spacecrafts and interference detection using aircraft based methods \cite{William1989}. However, these methods impose high computational burden and typically require long observation time. A potential method which overcome this limitations involve the use of movable spot beam antenna on domestic spacecrafts for locating unauthorized interference sources or minimizing the effects of interference on the actual satellite transmission. This require additional cost for spacecraft development and launching into orbit. Satellite based location methods use a single non-geostationary satellite to measure the direction of arrival of signals arriving from the interference sources. This can be achieved by using the relative movement information between the satellite and source of interference. Although this approach looks promising, it is not suitable for satellite in geostationary (GEO) orbit. 

Time difference of arrival (TDOA) interference location methods \cite{William1989, Chan2012} which compares the uplink propagation time  for two adjacent satellites in different orbital locations has been very popular. The fundamental idea is that since the relative position of the two satellites with respect to the ground receive station is known, the time difference of arrival between these satellites can be used to localize the uplink transmitter onto a curve on the Earth surface \cite{William1989}. TDOA dual satellite based TDOA method offer the advantage of locating interference sources regardless of their mission and without disruption of normal transmissions via the satellite. No additional spaceborne hardware is also required for their implementation. However, these methods require that the transponders on adjacent satellites be slightly occupied to prevent interference from each other.

Recently, interference location methods using single satellite have been considered in literature. These methods are based on the estimation of direction of arrivals using onboard multibeam antennas (MBA). Such method is the focus of \cite{Matsumoto1997}. These techniques, however, suffer from low direction estimation accuracy. In \cite{Tong2000}, a location estimation technique based on RBF neural network using satellite MBA model has been proposed. This offer a simple and computationally efficient method due to its learning ability and parallel processing and its also applicable to satellites in geostationary orbit. A similar approach based on support vector regression was also proposed in \cite{BaiHu2006}. This method consider the direction of arrival of the interference sources as a mapping from the MBA array space to the  space of arriving directions.

We present a subspace based method for interference location using MBA model in this paper, our method is based on the super-resolution and asymptotic maximum likelihood Multiple Signal Classification (MUSIC) which exploits the eigenstructure of the covariance matrix of the MBA signal model to extract information about the direction of arrival. 

The rest of the paper is structured as follows. Section \ref{sec:model} introduces the satellite multibeam antenna model upon which the direction estimation is based. In section \ref{sec:secMUSIC}, we present the proposed MUSIC based approach for interference location. Section \ref{sec:simulate} describes the simulation parameters along with a discussion of the results and comparison with existing methods. Finally, conclusion is drawn in section \ref{sec:secConc}.

\section{Satellite Multi-Beam Antenna Signal Model}
\label{sec:model}
\begin{figure}[t]
\centering
\includegraphics[width=1\columnwidth]{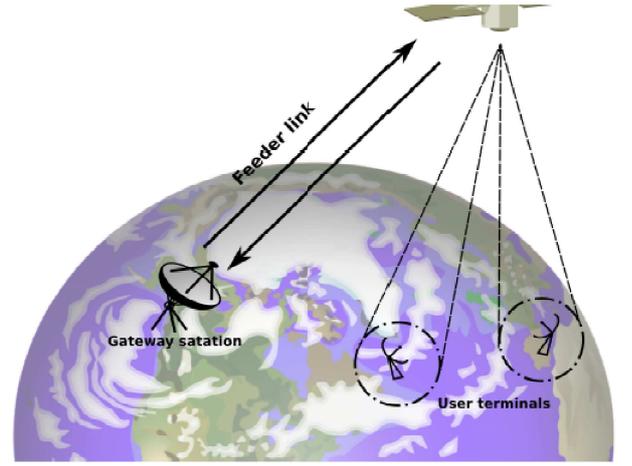}
\caption{Illustration of Satellite Systems with Multi-Beam Antenna Arrays \cite{Jesus2012}. A satellite with multibeam antennas communication with Earth stations. The transmission may be interfered by other undesired stations.}
\label{fig:fig1}
\end{figure}

This Section present the signal model for satellite systems with multibeam antenna array. Consider a satellite system with multibeam antenna (MBA) as shown in Figure~\ref{fig:fig1}. The multibeam antenna consist of several feeder that illuminate a single reflector to produce narrow spot beams directed in different spatial directions. The typical construction of satellite MBAs is such that the difference in path length of propagating wavefronts are equal between elements of the array. The propagating plane waves, therefore, have equal phases but different amplitude. We consider an MBA with M elemental spot beams as shown in Fig.~\ref{fig:fig2}. Similar to \cite{BaiHu2006}, we assume that $P$ narrowband signal sources with known centre frequency impinge on the multibeam antenna elements with directions of arrival
\begin{equation}
\label{eq:eq1}
\boldsymbol{\theta}=\left[\theta_1,\theta_2,\cdots,\theta_P\right]^T\in\mathbb{C}^{P\times 1}
\end{equation}
in the azimuth direction  and 
\begin{equation}
\label{eq:eq2}
\boldsymbol{\phi}=\left[\phi_1,\phi_2,\cdots,\phi_P\right]^T\in\mathbb{C}^{P\times 1}
\end{equation}
in the elevation spectrum. The received signal at the mth element of the multibeam array can be modelled as a superposition of the $P$ impinging waves as
\begin{equation}
\label{eq:eq3}
y_m(t)=\sum_{p=1}^{P}\alpha_m(\theta_p,\phi_p)s_p(t)+n_m(t)\quad\forall m=1,\cdots,M
\end{equation}
where $t$ is the time instant at which the observation is made, $s_p(t)$ is the transmitted signal from the pth interferer and $y_m(t)$ is the combined signal from the $P$ sources in noise. $n_m(t)$ is the received noise at the mth element. $\alpha_m(\theta_p,\phi_p)$ is the steering response of the mth element to the wave arriving with direction angles $\theta_p$ and $\phi_p$ in the azimuth and elevation domain. The steering response is defined as
\begin{equation}
\label{eq:eq4}
\alpha_m(\theta_p,\phi_p)=\sqrt{\eta}\frac{\pi DJ_1(\mu_{mp})}{\lambda\mu_{mp}}
\end{equation}
where $J_1(\cdot)$ denote the first order Bessel function, $\eta$ is a measure of the antenna efficiency and $D$ is the antenna aperture. $\lambda$ is the wavelength of the transmitted signal defined as
\begin{equation}
\label{eq:eq5}
\lambda = \frac{c}{2\pi f_c}
\end{equation}
with $c$ equal the velocity of light and $f_c$ is the center frequency. $\mu_{mp}$ is given as \cite{BaiHu2006}
\begin{equation}
\label{eq:eq6}
\mu_{mp}=\frac{\pi D}{\lambda}\sin\left(\sqrt{(\theta_p-\theta_m)^2+(\phi_p-\phi_m)^2}\right)
\end{equation}
where $\theta_m$ and $\phi_m$ are the azimuth and elevation angles of the $m$ beam spot center, respectively. Denoting 
\begin{equation}
\label{eq:eq7}
\mathbf{y}(t)=\left[y_1(t) \quad y_2(t)\quad\cdots\quad y_M(t)\right]^T\in\mathbb{C}^{M\times 1}
\end{equation}
A vector form for the data from all the beam spots is thus
\begin{equation}
\label{eq:eq8}
\mathbf{y}(t)=\boldsymbol{\alpha}(\boldsymbol{\theta},\boldsymbol{\phi})\mathbf{s}(t)+\mathbf{n}(t)
\end{equation}
where
\begin{align}
\mathbf{s}(t)&=\left[s_1(t)\quad\cdots\quad s_M(t)\right]^T\nonumber\\
\mathbf{n}(t)&=\left[n_1(t)\quad\cdots\quad n_M(t)\right]^T
\end{align}
The array steering matrix $\boldsymbol{\alpha}(\boldsymbol{\theta},\boldsymbol{\phi})$ is defined as
\begin{equation}
\label{eq:eq9}
\boldsymbol{\alpha}(\boldsymbol{\theta},\boldsymbol{\phi})=\begin{bmatrix}
\alpha_1(\theta_1,\phi_1) & \alpha_1(\theta_2,\phi_2)  & \cdot & \alpha_1(\theta_P,\phi_P)\\
\alpha_2(\theta_1,\phi_1) & \alpha_2(\theta_2,\phi_2)  & \cdot & \alpha_2(\theta_P,\phi_P)\\
\vdots & \vdots & \ddots & \vdots\\
\alpha_M(\theta_1,\phi_1) & \alpha_M(\theta_2,\phi_2)  & \cdot & \alpha_M(\theta_P,\phi_P)
\end{bmatrix}\in\mathbb{C}^{M\times P}
\end{equation}
Given the model in \eqref{eq:eq8}, the aim of the interference location scheme is to extract the $P$ parameter sets $[\theta_p,\phi_p]$ from the noisy observation data acquired using the multibeam antennas.
\section{DOA Estimation Using 2D MUSIC}\label{sec:secMUSIC}
In the previous section, we have described the data model for the satellite multibeam antenna observation upon which the DOA estimation in this section is based. The 2-dimensional MUSIC \cite{Schmidt} based estimation for interference location is presented in this section.
\subsection{Covariance Matrix Estimation}
Given the $K$ observations, $\mathbf{y}(1),\mathbf{y}(2),\cdots,\mathbf{y}(K)$, we form the data matrix as
\begin{equation}
\label{eq:eq10}
\mathbf{Y}=\left[\mathbf{y}(1)\quad\mathbf{y}(2)\quad\cdots\quad\mathbf{y}(K)\right]\in\mathbf{M\times K}
\end{equation}
The spatial covariance matrix is then estimated using\footnote{Note that in practice, data preprocessing methods for reducing or eliminating noise may be applied before the actual estimation.}
\begin{equation}
\label{eq:eq11}
\mathbf{R}_{yy}=\frac{X*X^\dagger}{K}\in\mathbb{C}^{M\times M}
\end{equation}
where $[\cdot]^\dagger$ denote the Hermitian conjugate transpose.
\subsection{Subspace Decomposition}
Using the model in \eqref{eq:eq8}, the spatial covariance matrix can be shown to be \cite{Kay1}
\begin{align}
\label{eq:eq12}
\mathbf{R}_{yy}&=\mathbf{E}[\mathbf{y}\mathbf{y}^\dagger]\nonumber\\
&=\boldsymbol{\alpha}(\boldsymbol{\theta},\boldsymbol{\phi})\mathbb{E}[\mathbf{s}\mathbf{s}^\dagger]\boldsymbol{\alpha}(\boldsymbol{\theta},\boldsymbol{\phi})^\dagger+\mathbb{E}[\mathbf{n}\mathbf{n}^\dagger]\nonumber\\
&=\boldsymbol{\alpha}(\boldsymbol{\theta},\boldsymbol{\phi})\mathbf{R}_{ss}\boldsymbol{\alpha}(\boldsymbol{\theta},\boldsymbol{\phi})^\dagger+\mathbb{E}[\mathbf{n}\mathbf{n}^\dagger]
\end{align}
where $\mathbf{R}_{ss}=\mathbb{E}[\mathbf{s}\mathbf{s}^\dagger]$ is the covariance matrix of the transmitted signals from the interference sources. Assuming that the received noise is Gaussian with variance $\sigma^2$. \eqref{eq:eq12} can then be expressed as
\begin{equation}
\label{eq:eq13}
\mathbf{R}_yy=\boldsymbol{\alpha}(\boldsymbol{\theta},\boldsymbol{\phi})\mathbf{R}_{ss}\boldsymbol{\alpha}(\boldsymbol{\theta},\boldsymbol{\phi})^\dagger+\sigma^2\mathbf{I}
\end{equation}
where $\mathbf{I}$ is the $M\times M$ identity matrix. The eigendecomposition of $\mathbf{R}_{yy}$ is defined as
\begin{align}
\mathbf{R}_{yy}&=\left[\mathbf{U}_s\quad \mathbf{U}_n\right]\begin{bmatrix}
\Lambda_s & \quad\\ \quad & \Lambda_n
\end{bmatrix}\begin{bmatrix}
\mathbf{U}_s^\dagger\\ \mathbf{U}_n^\dagger
\end{bmatrix}\\
&=\mathbf{U}_s\Lambda_s\mathbf{U}_s^\dagger+\mathbf{U}_n\Lambda_n\mathbf{U}_n^\dagger
\end{align}
where $\mathbf{U}_s$ and $\mathbf{U}_n$ are the signal and noise subspace eigenvectors, respectively. The corresponding eigenvalues are contained in the diagonal matrices $\Lambda_s$ and $\Lambda_n$.
\begin{figure}[!t]
\centering
\includegraphics[width=4in]{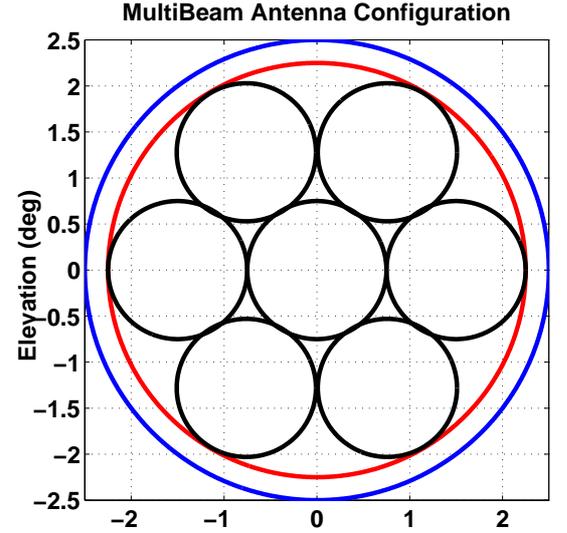}
\caption{A typical satellite multibeam antenna configuration showing seven feeders (spot beams) in an hexagonal lattice structure. The 3dB beam width of each feed is about $1.5^0$. A generalized structure with M spot beams will be considered in this paper. }
\label{fig:fig2}
\end{figure}
\subsection{MUSIC Pseudospectrum}
Since the array steering vector in the direction of an interferer is orthogonal to the noise subspace, the product $\mathbf{a}^\dagger(\theta,\phi)\mathbf{U}_n\mathbf{U}_n^\dagger\mathbf{a}(\theta,\phi)=0$ when the angles corresponds to the DOA of an interference transmitting source. The direction of interferers can therefore be estimated as the $P$ largest peaks of 
\begin{equation}
\label{eq:eq14}
\mathbf{P}_M(\theta,\phi)=\frac{1}{\mathbf{a}^\dagger(\theta,\phi)\mathbf{U}_n\mathbf{U}_n^\dagger\mathbf{a}(\theta,\phi)}
\end{equation}
where $P_M$ is the 2-D MUSIC pseudospectrum. An alternative representation for \eqref{eq:eq14} is
\begin{equation}
\label{eq:eq15}
\mathbf{P}_M(\theta,\phi)=\frac{\mathbf{a}^\dagger(\theta,\phi)\mathbf{a}(\theta,\phi)}{\mathbf{a}^\dagger(\theta,\phi)\mathbf{U}_n\mathbf{U}_n^\dagger\mathbf{a}(\theta,\phi)}
\end{equation}

\section{Numerical Simulations}\label{sec:simulate}
\begin{figure}[h]
\centering
\includegraphics[height=2.4in,width=1\columnwidth]{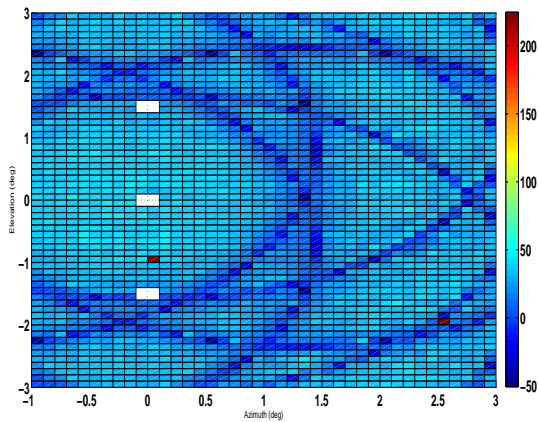}
\caption{2D Plot of MUSIC Pseudospectrum for DOA Estimation Using Satellite MultiBeam Antennas at SNR $=$20dB. Two interference sources with angles $[-1.0, 2]$ and $[-2.0, 2.5]$.}
\label{fig:fig3}
\end{figure}

\begin{figure}[h]
\centering
\includegraphics[height=2.4in,width=1\columnwidth]{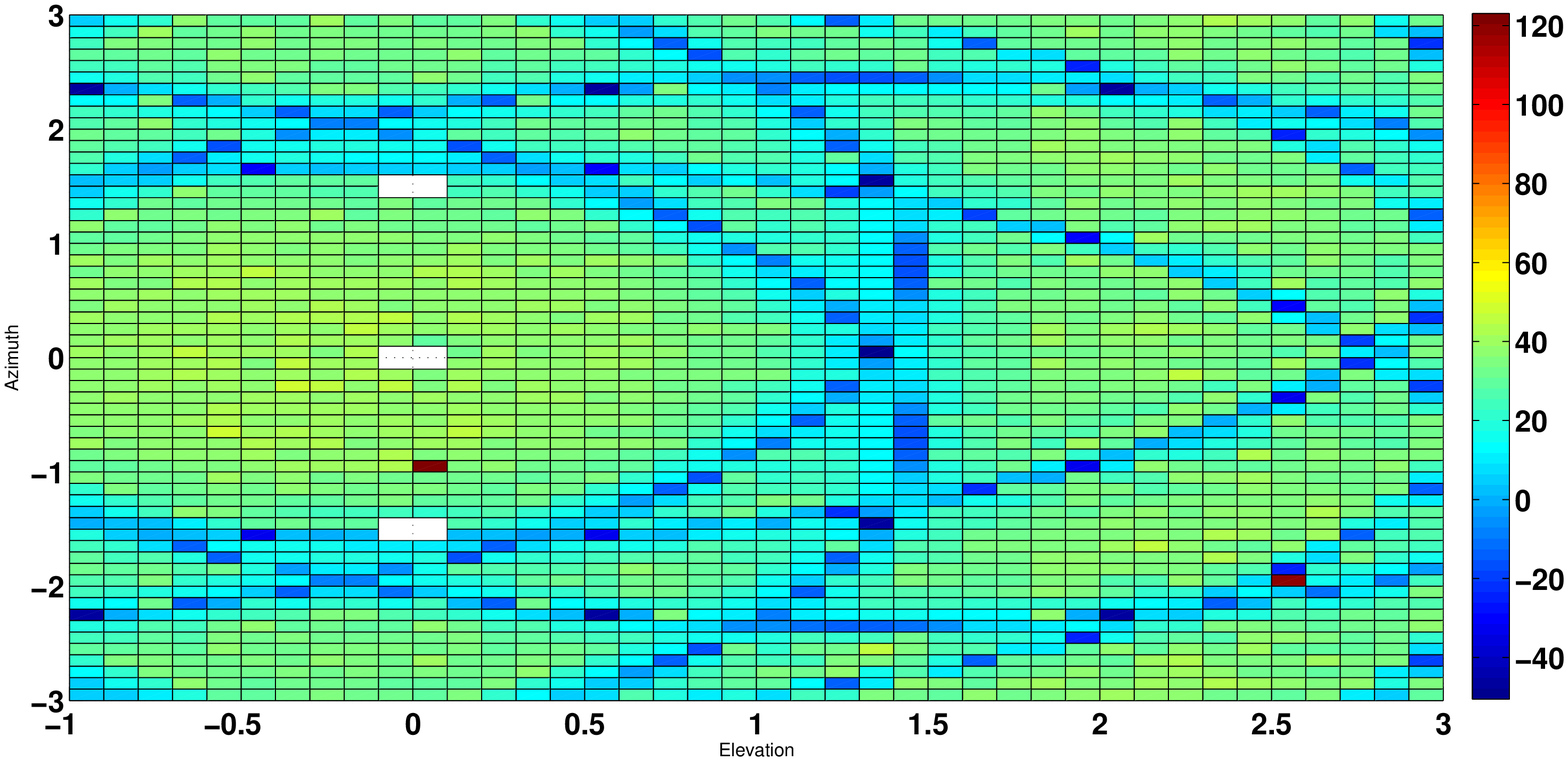}
\caption{2D Plot of MUSIC Pseudospectrum for DOA Estimation Using Satellite MultiBeam Antennas at SNR $=$10dB. Two interference sources with angles $[-1.0, 2]$ and $[-2.0, 2.5]$.}
\label{fig:fig4}
\end{figure}

\begin{figure}[h]
\centering
\includegraphics[height=2.4in,width=1\columnwidth]{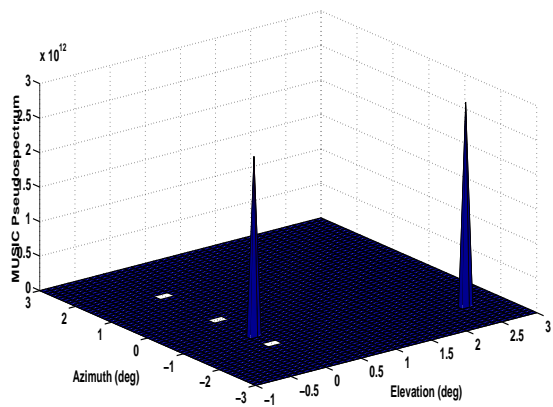}
\caption{Plot of MUSIC Pseudospectrum for DOA Estimation Using Satellite MultiBeam Antennas at SNR $=$10dB. Two interference sources with angles $[-1.0, 2]$ and $[-2.0, 2.5]$. It illustrates that MUSIC exhibits peaks at locations corresponding to the interference sources.}
\label{fig:fig5}
\end{figure}

\begin{figure}[h]
\centering
\includegraphics[height=2.4in,width=1\columnwidth]{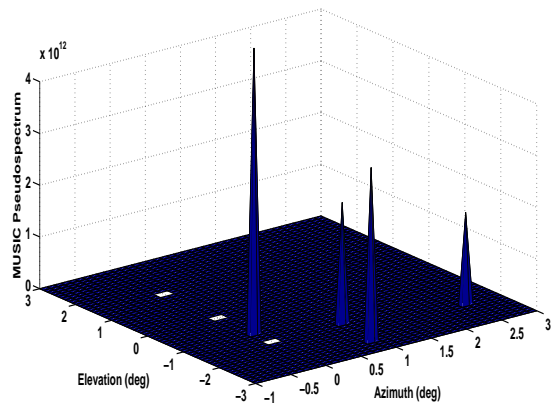}
\caption{MUSIC Pseudospectrum with four interference sources at SNR$=$10dB.}
\label{fig:fig6}
\end{figure}

\begin{figure}[h]
\centering
\includegraphics[height=2.4in,width=1\columnwidth]{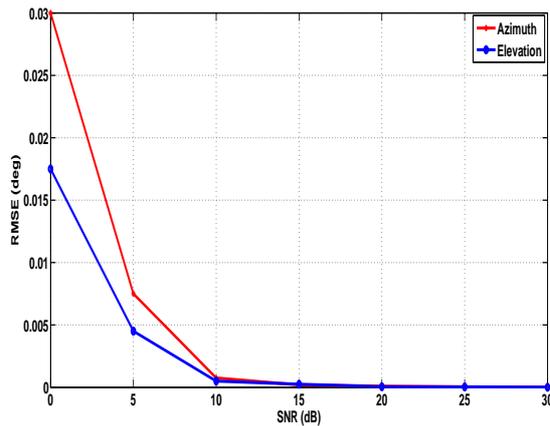}
\caption{Averaged RMSE versus SNR of the proposed algorithm for both the azimuth and elevation interference directions. It shows that the estimation accuracy increases with increasing SNR.}
\label{fig:fig7}
\end{figure}

\begin{figure}[h]
\centering
\includegraphics[height=2.4in,width=1\columnwidth]{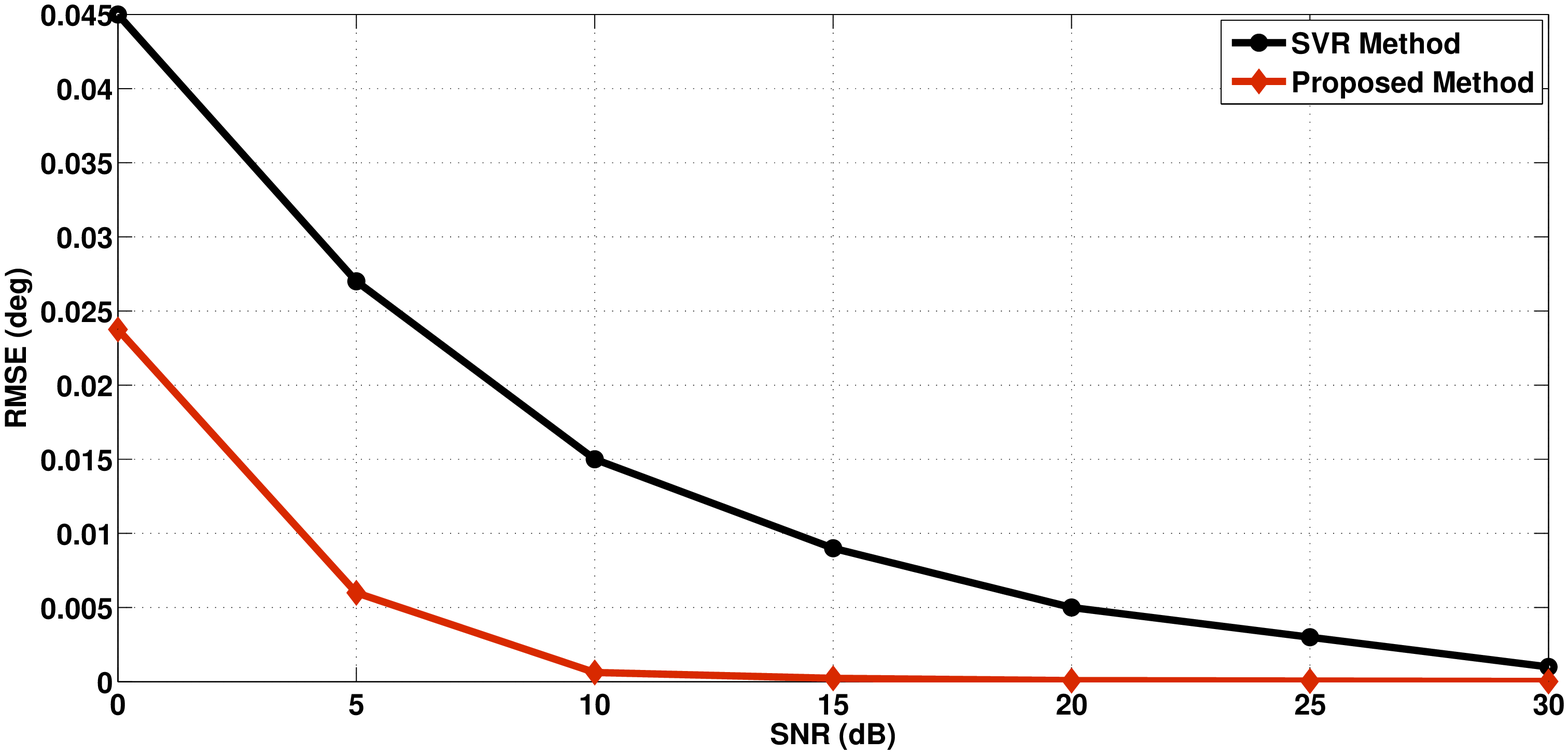}
\caption{Averaged RMSE versus SNR for the proposed scheme and the support vector regression (SVR) based interference location method. The RMSE is averaged over the sources and all the elevation and azimuth domain. We observe that our algorithm outperforms the SVR method at all SNR values.}
\label{fig:fig8}
\end{figure}

In this section, we evaluate the performance of the proposed location scheme\footnote{The MATLAB implementation of the MUSIC based approach is given in Appendix A. This is to allow reproducibility of our research results and experimentation of the proposed scheme.} and compare with the SVR and RBF neural network methods. The algorithms are evaluated in terms of root mean square error defined as
\begin{equation}
RMSE(\theta_p)=\sqrt{\frac{\sum_{c=1}^C(\theta_p-\hat{\theta}_p)^2}{C}}
\end{equation}
and
\begin{equation}
RMSE(\phi_p)=\sqrt{\frac{\sum_{c=1}^C(\phi_p-\hat{\phi}_p)^2}{C}}
\end{equation}
where $\hat{G}$ denote the estimated value of $G$ and $C$ is the number of Monte-Carlo simulations. The averaging is performed in this paper using 1000 independent runs, i.e $C=1000$. We consider a satellite multibeam antenna configuration as shown in Figure~\ref{fig:fig2} with $M=7$ beam spot (feeders) in an hexagonal lattice structure. We consider two different scenarios in our experiment. The first scenario is a simplified case with only two interfering signal sources and the second scenario has four interferers. We set the azimuth/elevation angles of the interfering radio sources to $[-1.0, 2]$ and $[-2.0, 2.5]$ in the two source scenario. In the four interferers case, we additionally introduce $[-1.5,1]$ and $[-2.5, 0.9]$. In Fig.~\ref{fig:fig3}, we present a 2D plot of the MUSIC pseudo-spectrum for the two sources case at a signal to noise (SNR) value of $20\,$dB. It shows that the proposed scheme can accurately detect the azimuth and elevation directions of the two interference sources.  A similar plot obtained at a lower SNR value of $10\,$dB is shown in Fig.~\ref{fig:fig4}. As can be seen from the figure, the two interference sources are clearly detected. This shows the high resolution performance of the proposed scheme. In Fig.~\ref{fig:fig5}, we plot the MUSIC psedospectrum in three dimensions showing the azimuth and elevation angles and the power of the detected sources. It shows that the MUSIC spectrum exhibit very large peaks where the angles corresponds to the direction of the interferers. Figure~\ref{fig:fig6} shows the 3D plot of the Pseudospectrum for the four sources case. We also observe that the algorithm can detect with high accuracy all the four interference signal sources. Fig.~\ref{fig:fig7} shows the root mean square error (RMSE) of the proposed algorithm as a function of SNR. We observe that the estimation performance improves with increasing SNR for both the azimuth and elevation angles. Note that the difference in RMSE for azimuth and elevation angles is likely due to the difference in magnitude of the actual directions. In Fig.~\ref{fig:fig8}, we plot the RMSE versus SNR for the MUSIC based algorithm and the support vector regression (SVR) method \cite{BaiHu2006}. The figure shows that the algorithm outperform the SVR method at all SNR values. 

\section{Conclusion}\label{sec:secConc}
This paper investigates single satellite interference location methods for communication satellite in geostationary and non-geostationary orbits. A MUSIC based subspace method is proposed for the estimation of the interference location using data from multibeam antennas onboard the satellite. The performance of the proposed algorithm is analyzed and comparison is made with the RBF neural network and support vector regression (SVR) methods using the root mean square error criterion. Simulation results show that the proposed method offer improved location estimation performance compared with previous method with a slight increase in computational complexity. Future work will investigate interference location using subspace methods that eliminate the peak search requirement.

\bibliographystyle{ijcaArticle}
\bibliography{SatelliteBiB}

\section*{Author Biography}
\parpic{\includegraphics[width=1in,clip,keepaspectratio]{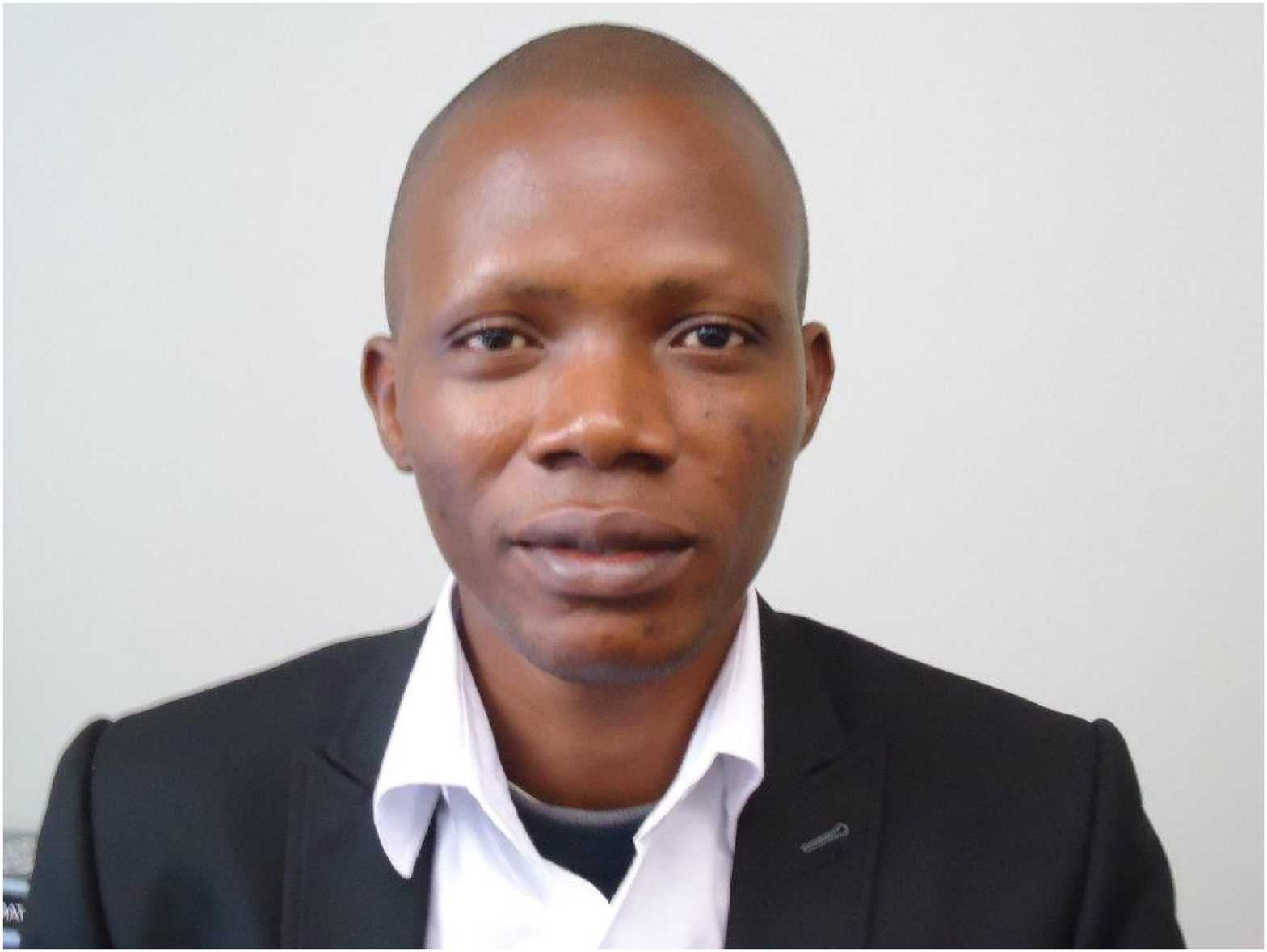}}
\emph{\bf Ramoni Adeogun}  is currently working towards a PhD degree in Engineering (specializing in Communications and Signal Processing) at the School of Engineering and Computer Science, Victoria University of Wellington, New Zealand. He received the B.Eng degree in Electrical and Computer Engineering with First Class Honours from the Federal University of Technology Minna, Niger State Nigeria in 2007. Between 2008 - 2009, he was with the Information and Communication Technology (ICT) directorate, University of Jos, Nigeria. He worked briefly as an Engineer with Odua Telecoms Ltd, Ibadan Nigeria in 2009. He joined the National Space Research and Development Agency (NASRDA) Abuja Nigeria in 2010 and has since been working with the Engineering and Space Systems (ESS) division of the agency. Ramoni holds several Honours and awards including Ogun State tertiary Scholarship (2003 -2006), best graduating student in the university (2007), Commonwealth Shared Scholarship (2011) and Victoria Doctoral Scholarship. He is a graduate member of Institute of Electrical and Electronics Engineers. A member of the International Association of Engineers (IAENG).
\clearpage
\appendix
\section{MATLAB CODE}
\begin{verbatim}
%MUSIC based Satellite Interference Location Method
%Estimates Azimuth and Elevation Direction from 
%Multi-Beam Satellite Antenna Array
%By R.O Adeogun - Oct. 2013
%============================================%
               %House Keeping
%============================================%
close all
clear all
clc
%===================Simulation Parameters=============
N = 7;                             %Number of antennas
M = 2;                            %Number of interference
ThetaC = [ -1.5 0 1.5 -0.75 0.75 -0.75 0.75];
PhiC = [0 0 0 1.25 1.25 -1.75 -1.75];

ThetaI = [ -1.0 -2.0];
PhiI = [0 2.5];
eta = 0.1;
K = 1000;
D = 10;
c = 3e8;
f = 2.9e9;
Lambda = c/(2*pi*f);
SNR=100;                             %SNR in decibel
SNR_linear=10.^(-SNR./10);

%==MBA Array Steering Steering=========================
for i = 1:M
    T1 = ThetaI(i)-ThetaC;
    P1 = PhiI(i)-PhiC;
    u1=(pi*D)/Lambda*sin((sqrt(T1.^2+P1.^2)));
    J1 = besselj(0,u1);
    A(:,i) = sqrt(eta)*(pi*D)/Lambda*J1./u1;
end

%=====================Data Matrix=================
for k =1:K
    s = 1/sqrt(2)*(randn(1,M)+1j*randn(1,M));
    X(:,k) = A*s.';
end

%==================Add Background Noise===============
X = X+1/sqrt(2)*SNR_linear*(randn(N,K)+1j*randn(N,K));

%===============Covariance Matrix Estimation=======
R = X*X'/K;

%==============EVD and Subspace Decomposition=======
[W B]= eig(R);
[DD S]=sort(diag(B),'descend');
Es=W(:,S(1:M));
En=W(:,S(M+1:end));


%=================MUSIC Pseudospectrum==========
ThetaS = [-3:0.1:3];
PhiS = [-1:0.1:3];

for ii=1:length(ThetaS)
    for jj = 1:length(PhiS)
        T1 = ThetaS(ii)-ThetaC;
        P1 = PhiS(jj)-PhiC;
        u1=(pi*D)/Lambda*sin((sqrt(T1.^2+P1.^2)));
        J1 = besselj(0,u1);
        an = sqrt(eta)*(pi*D)/Lambda*J1./u1;
        Pmusic(ii,jj)=1./abs(an*En*En'*an');
    end
end

%=================Plotting==========================
figure(1)
surf(PhiS,ThetaS,10*log10(Pmusic))
\end{verbatim}
\end{document}